\documentclass[12pt,twoside]{article}
\usepackage[mathscr]{eucal}
\usepackage{amsmath,amsfonts,amssymb,amsthm}
\usepackage[matrix,arrow]{xy}
\usepackage{times}
\usepackage{pdfsync}
\usepackage{graphicx}
\usepackage{epstopdf}
\usepackage{dcolumn}
\usepackage{hyperref}
\usepackage{color}

\def\d_Vphi{\text{d}_V\hspace{-0.06em}\phi}
\def\d_Vphibar{\text{d}_V\hspace{-0.06em}\bar\phi}
\def\d_Vxi{\text{d}_V\hspace{-0.06em}\xi}

\def\be{\begin{eqnarray}}
\def\ee{\end{eqnarray}}
\def\beann{\begin{eqnarray*}}
\def\eeann{\end{eqnarray*}}
\def\beq{\begin{equation}}
\def\eeq{\end{equation}}
\def\ba{\begin{array}}
\def\ea{\end{array}}
\def\ben{\begin{enumerate}}
\def\een{\end{enumerate}}
\def\bea{\begin{eqnarray}}
\def\eea{\end{eqnarray}}

\def\5{\bar }
\def\6{\partial }
\def\7{\hat }
\def\4{\tilde }
\advance\textheight\topskip
\renewcommand{\tilde}{\widetilde}
\renewcommand{\hat}{\widehat}
\newtheorem{prop}{Proposition}[section]

\newtheorem{theorem}[prop]{Theorem}

\renewcommand{\d}{\partial}

\newcommand{\binner}[2]{%
  {\langle}\kern-4.15pt{\langle}#1{,}\,#2{\rangle}\kern-4.15pt{\rangle}}

\newcommand{\half}{\frac{1}{2}}

\newcommand{\ffrac}[2]{\raisebox{.5pt}%
  {\footnotesize$\displaystyle\frac{#1}{#2}$}\kern1pt}

\def\cL{\mathcal{L}}

\def\cX{\mathcal{X}}
\def\cY{\mathcal{Y}}

\numberwithin{equation}{section} \makeatletter
\@addtoreset{equation}{section}

\DeclareFontFamily{OT1}{rsfs}{} \DeclareFontShape{OT1}{rsfs}{m}{n}{
<-7> rsfs5 <7-10> rsfs7 <10-> rsfs10}{}
\DeclareMathAlphabet{\mycal}{OT1}{rsfs}{m}{n}

\begin{document}
\

\def\mytitle{Asymptotic symmetries at null infinity\\ and local conformal properties of spin coefficients}

\pagestyle{myheadings} \markboth{\textsc{\small}}{%
  \textsc{\small}}
\addtolength{\headsep}{4pt}

\begin{flushright}
\small ULB-TH/13-02
\end{flushright}


\begin{centering}

  \vspace{1cm}

  \textbf{\Large{\mytitle}}



  \vspace{1.5cm}

  {\large Glenn Barnich$^{a}$ and Pierre-Henry Lambert$^{b}$}

\vspace{.5cm}

\begin{minipage}{.9\textwidth}\small \it \begin{center}
   Physique Th\'eorique et Math\'ematique\\ Universit\'e Libre de
   Bruxelles\\ and \\ International Solvay Institutes \\ Campus
   Plaine C.P. 231, B-1050 Bruxelles, Belgium \end{center}
\end{minipage}

\end{centering}

\vspace{1cm}

\begin{center}
  \begin{minipage}{.9\textwidth}
    \textsc{Abstract}.   We show that the symmetry algebra of asymptotically flat four
  dimensional spacetimes at null infinity in the sense of Newman and
  Unti is isomorphic to the direct sum of the abelian algebra of
  infinitesimal conformal rescalings with $\mathfrak{bms}_4$. We then
  work out the local conformal properties of the relevant
  Newman-Penrose coefficients, as well as the surface charges and
  their algebra.
  \end{minipage}
\end{center}

\vfill

\textit{Proceedings of the} \textbf{International Conference on Quantum Field Theory and Gravity}\textit{, July 31--August 4, 2012, Center of Theoretical Physics, Tomsk State Pedagogical University, Russia, the} \textbf{Workshop on Gravity, Quantum, and Black Holes} \textit{at the International Conference on Mathematical Modeling in Physical Sciences, September 3--7, 2012, Budapest, Hungary, and the \textbf{8th edition of the Workshop on Quantum Field Theory and Hamiltonian Systems}, September 19-22, 2012, Craiova, Romania.}

\noindent
\mbox{}
\raisebox{-3\baselineskip}{%
  \parbox{\textwidth}{\mbox{}\hrulefill\\[-4pt]}}
{\scriptsize$^a$Research Director of the Fund for Scientific
  Research-FNRS Belgium. E-mail: gbarnich@ulb.ac.be\\$^b$
Boursier ULB. E-mail: Pierre-Henry.Lambert@ulb.ac.be}

\thispagestyle{empty}
\newpage


\section{Introduction}

This conference proceedings summarizes the results of paper
\cite{Barnich:2011nu} to which we refer for detailed computations and
discussions.

The definitions of asymptotically flat four dimensional space-times at
null infinity by Bondi-Van der Burg-Metzner-Sachs
\cite{Bondi:1962px,Sachs:1962wk} (BMS) and Newman-Unti (NU)
\cite{newman:891} in 1962 merely differ by the choice of the radial
coordinate. Such a change of gauge should not affect the asymptotic
symmetry algebra if, as we contend, this concept is to have a major
physical significance. The problem of comparing the symmetry algebra
in both cases is that, besides the difference in gauge, the very
definitions of these algebras are not the same. Indeed, NU allow the
leading part of the metric induced on Scri to undergo a conformal
rescaling. When this generalization is considered in the BMS setting,
it turns out that the symmetry algebra is the direct sum of the BMS
algebra $\mathfrak{bms}_4$ \cite{Sachs2} with the abelian algebra of
infinitesimal conformal rescalings \cite{Barnich:2009se},
\cite{Barnich:2010eb}.

In this note we show that, as expected, the asymptotic symmetry algebra in the NU framework is again the direct sum of $\mathfrak{bms}_4$
with the abelian algebra of infinitesimal conformal rescalings of the
metric on Scri and thus coincides, as it should, with the generalized
symmetry algebra in the BMS approach.
We then discuss the transformation properties of the
Newman-Penrose coefficients parametrizing solution space in the NU
approach, focussing on the inhomogeneous terms in the
transformation laws that contain the information on the central
extensions of the theory, and we finally study the associated surface
charges and their algebra by following the analysis in the BMS gauge
\cite{Barnich:2011mi}.

\section{NU metric ansatz and asymptotic symmetries}
\label{sec:from-bondi-van}

The metric ansatz of NU can be written as
\begin{equation}
  \label{eq:10}
  ds^2=Wdu^2-2dr du+
g_{AB}(dx^A-V^Adu)(dx^B-V^Bdu)\,,
\end{equation}
with coordinates $u,r,x^A$ and
where
$
g_{AB}dx^Adx^B=r^2\bar\gamma_{AB}dx^Adx^B
+r C_{AB}dx^Adx^B+o(r)\,,\label{eq:11}
$
with $\bar\gamma_{AB}$ conformally flat. Below, we will use standard
stereographic coordinates $\zeta=\cot{\frac{\theta}{2}}e^{i\phi},\bar
\zeta$, $\bar\gamma_{AB}dx^Adx^B=e^{2{\tilde\varphi}} d\zeta
d\bar\zeta$, ${\tilde\varphi}={\tilde\varphi}(u,x)$.
There is also an additional condition, related to the fixing of the origin of the affine parameter of the null geodesic generators of the null hypersurfaces used to build the metric \cite{newman:891}, which yields here $C^A_A=0$ \cite{Barnich:2011nu}.

In the following we denote by $\bar D_A$ the covariant derivative with
respect to $\bar \gamma_{AB}$ and by $\bar\Delta$ the associated
Laplacian. The fall-off conditions are
$
V^A=O(r^{-2})$ and $W=-2 r\d_u
{\tilde\varphi}+\bar\Delta {\tilde\varphi}+O(r^{-1})\,,
$
where $\bar \Delta
{\tilde\varphi}=4 e^{-2{\tilde\varphi}}\d\bar\d{\tilde\varphi}$ with $\d=
\d_\zeta,\bar\d=\d_{\bar\zeta}$.


The infinitesimal NU transformations are defined as those
infinitesimal transformations that leave the form of the metric and
the fall-off conditions invariant, up to a
rescaling of the conformal factor
$\delta{\tilde\varphi}(u,x^A)={\tilde\omega}(u,x^A)$, and are in this case generated by
\begin{gather}
  \label{eq:26}
\left\{\begin{array}{l}
  \xi^u=f,\\
\xi^A=Y^A+I^A, \quad  I^A=- \d_B f\int_r^\infty dr^\prime
g^{AB},\\
\xi^r=-r \d_u f+Z+ J,\quad J=\d_A f \int_r^\infty dr^\prime V^A,
\end{array}\right.
\end{gather}
with $\d_r f=0=\d_r Y^A=\d_r Z$, $Z=\half \bar\Delta f$, $\d_u Y^A=0$, with $Y^A$ a conformal Killing vector of $\bar\gamma_{AB}$,
i.e. $Y^\zeta \equiv Y=Y(\zeta),\quad  Y^{\bar \zeta} \equiv \bar Y=\bar
  Y(\bar \zeta)$
in the coordinates ($\zeta,\bar\zeta$), and also with
\begin{equation}
  \label{eq:14}
  f=e^{
    {\tilde\varphi}}\big[ \tilde T+
  \half\int_0^u du^\prime
  e^{- {\tilde\varphi}}\tilde \psi\big],\quad  \tilde T= \tilde T(\zeta,\bar \zeta),
\end{equation}
with $\psi=\bar D_AY^A$ and $\tilde\psi=\psi-2\tilde\omega$.
Asymptotic Killing vectors thus depend on $Y^A,\tilde
T,{\tilde\omega}$ and the metric, $\xi=\xi[Y,\tilde
T,{\tilde\omega};g]$.  For such metric dependent vector fields,
consider the suitably modified Lie bracket taking the metric
dependence of the spacetime vectors into account, $
[\xi_1,\xi_2]_M=[\xi_1,\xi_2]-\delta^g_{ \xi_1}\xi_2+\delta^g_{
  \xi_2}\xi_1, $ where $\delta^g_{\xi_1}\xi_2$ denotes the variation
in $\xi_2$ under the variation of the metric induced by $\xi_1$,
$\delta^g_{ \xi_1}g_{\mu\nu}=\cL_{\xi_1}g_{\mu\nu}$.  Consider now the
extended $\mathfrak{bms}_4$ algebra, i.e., the semi-direct sum of the
algebra of conformal Killing vectors of the Riemann sphere with the
abelian ideal of infinitesimal supertranslations, trivially extended
by infinitesimal conformal rescalings of the conformally flat
degenerate metric on Scri. The commutation relations are given by
$[(Y_1,\tilde T_1,{\tilde\omega}_1),(Y_2,\tilde
T_2,{\tilde\omega}_2)]=(\hat Y,\hat {\tilde T},\hat{{\tilde\omega}})$
where
\begin{equation}
  \left\{\begin{array}{l}
      \label{eq:5}\hat Y^A= Y^B_1\d_B
Y^A_2-Y^B_2\d_B Y^A_1,\\
\hat {\tilde T}=Y^A_1\d_A
\tilde  T_2-Y^A_2\d_A \tilde T_1 +\half (\tilde T_1\d_AY^A_2-\tilde T_2\d_AY^A_1),\\
\hat{{\tilde\omega}}=0\,.
\end{array}\right.
\end{equation}
In these terms,  one can show the following:  
\begin{theorem}
  The spacetime vectors $\xi[Y,\tilde T,{\tilde\omega};g]$ realize the extended
  $\mathfrak{bms}_4$ algebra in the modified Lie bracket,
\begin{equation}
  \Big[\xi[Y_1,\tilde T_1,{\tilde\omega}_1;g],\xi[Y_2,\tilde T_2,{\tilde\omega}_2;g]\Big]_M=
\xi[\hat Y,\hat {\tilde T},\hat{{\tilde\omega}};g]\,,\label{eq:1}
\end{equation}
in the bulk of an asymptotically flat spacetime in the sense of Newman
and Unti.
\end{theorem}
Note in particular that for two different choices of the conformal
factor ${\tilde\varphi}$ which is held fixed, ${\tilde\omega}=0$, the
asymptotic symmetry algebras are isomorphic to $\mathfrak{bms}_4$,
which is thus a gauge invariant statement.

\section{Explicit relations between the NU and the BMS gauges\\ and local conformal transformation laws of the NU coefficients}
~ The choice of the radial coordinate in the definition of
asymptotically flat space-times in the BMS \cite{Bondi:1962px},
\cite{Sachs:1962wk}, \cite{Sachs2} and the NU \cite{newman:891}
approaches differs but the relation between the two radial coordinates
does not involve constant terms \cite{Barnich:2011nu} and is of the
form $ r^\prime=r +O(r^{-1})\label{eq:3}\,.  $ This change of
coordinates only affects lower order terms in the asymptotic expansion
of the metric that play no role in the definition of asymptotic
symmetries and explains a posteriori why the asymptotic symmetry
algebras in both approaches are isomorphic.

In the BMS set-up, the general solution to Einstein's field equations
is parametrized by some functions \cite{Bondi:1962px},
\cite{Sachs:1962wk}, \cite{Barnich:2010eb} among which are the mass
and angular momentum aspects, and the news tensors. In the NU case
instead \cite{newman:891}, the free data characterizing solution space
are described in terms of the spin coefficient $\sigma^0$ and its time
derivative, and also in terms of the $\Psi_\alpha^0$ (with
$\alpha=0,1,2,3,4$), five complex scalars representing all the
components of the Weyl tensor.  The explicit relations between the
free data characterizing asymptotic solution space in both approaches
were established for instance in \cite {Barnich:2011nu}.

Using the ``eth'' operators \cite{newman:863} defined for a field
$\eta^s$ of spin weight $s$ according to the conventions of
\cite{Penrose:1986} through $ \eth \eta^s= P^{1-s}\bar
\d(P^s\eta^s)\,, \bar \eth \eta^s=P^{1+s}\d(P^{-s}\eta^s)$ with $
P=\sqrt 2 e^{-\tilde\varphi}\,, $ where $\eth,\bar\eth$ raise
respectively lower the spin weight by one unit
and let $\cY=P^{-1} \bar Y$ and $\bar \cY=P^{-1}
Y$. The conformal Killing equations and the conformal factor then
become
$\eth \bar \cY=0=\bar\eth \cY$ and $\psi=(\eth \cY+\bar\eth
  \bar \cY)$.
Using the notation $S=(Y,\tilde T,\tilde \omega)$, we have $-\delta_S
\bar \gamma_{AB}=2\tilde\omega \bar\gamma_{AB}$ for the background
metric.

To work out the transformation properties of the NU coefficients
characterizing asymptotic solution space, one needs to evaluate the
subleading terms in the Lie derivative of the metric on-shell. This
can also be done by translating the results from the BMS gauge, using
the dictionary of \cite{Barnich:2011nu}, which yields in this case
\begin{equation}
\begin{split}
  \label{eq:16b}
  -\delta_S \sigma^0 & = [f\d_u+\cY\eth+ \bar
  \cY\bar\eth+\frac{3}{2}\eth \cY-\frac{1}{2} \bar\eth \bar \cY-\tilde
  \omega] \sigma^0-\eth^2
  f\,,\\
  -\delta_S \dot\sigma^0 & = [f\d_u+ \cY\eth + \bar \cY\bar\eth+2\eth
  \cY-2\tilde\omega]\dot\sigma^0-\half \eth^2\tilde \psi\,,\\
-\delta_S\Psi^0_i&=[f\d_u+\cY\eth+\bar\cY\bar\eth+\frac{5-i}{2}\eth\cY
+\frac{1+i}{2}\bar\eth\bar\cY-3\tilde\omega]\Psi^0_i+(4-i)\eth f\Psi^0_{i+1}\,,\\
\end{split}
\end{equation}
with $i=1,2,3,4$.

\section{Surface charge algebra}
\label{sec:surf-charge-algebra}

In this section, $\tilde\omega=0$ so that $f=T+\half u\psi$ and we use the
notation $s=(\cY,\bar\cY,T)$ for elements of the symmetry algebra,
which is given in these terms by $[s_1,s_2]=\hat s$ where 
\begin{equation}
\begin{gathered}
  \label{eq:57}
  \hat \cY=\cY_1\eth \cY_2 -(1\leftrightarrow 2),\qquad  \hat
  {\bar\cY}=\bar \cY_1\bar \eth \bar \cY_2 -(1\leftrightarrow 2),\\
  \hat T= (\cY_1\eth +\bar \cY_1\bar \eth)T_2-\half \psi_1 T_2 -(1\leftrightarrow 2)\,.
\end{gathered}
\end{equation}
The translation of the charges, the non-integrable piece due to the
news and the central charges computed in \cite{Barnich:2011mi} gives
here
\begin{flalign}
  Q_{s}[\cX]&=-\frac{1}{8\pi G}\int d^2\Omega^\varphi \Big[\big(
  f(\Psi^0_2+\sigma^0\dot{\bar
    \sigma}^0 )+\cY(\Psi^0_1
  +\sigma^0\eth\bar\sigma^0+\half\eth(\sigma^0\bar\sigma^0))\big)
  +{\rm c.c.}\Big],\nonumber\\
  \Theta_{s}[\delta\cX,\cX]&=\frac{1}{8 \pi G}\int d^2
  \Omega^\varphi\, f \big[\dot{\bar\sigma}^0\delta\sigma^0 +{\rm
    c.c.}\big]\,,  \label{eq:19}\\
  K_{s_1,s_2}[\cX]&= \frac{1}{8 \pi G}\int d^2 \Omega^\varphi\, \Big[
  \big(\frac{1}{4} f_1 \eth f_2 \bar\eth \bar R+\half \bar\sigma^0 f_1 \eth^2
  \psi_2 - (1\leftrightarrow 2) \big) + {\rm c.c.} \Big]\,.\nonumber
\end{flalign}

We recognize all the ingredients of the surface charges described in
\cite{0264-9381-1-1-005}. More precisely the angular
(super-)momentum that we get is 
\begin{equation}
  Q_{\cY,0,0}=-\frac{1}{8\pi G}\int d^2\Omega^\varphi\,
  \cY\Big[\Psi^0_1
  +\sigma^0\eth\bar\sigma^0+\half\eth(\sigma^0\bar\sigma^0)
  - \frac{u}{2}\eth\big(\Psi^0_2+\bar\Psi^0_2+
  \d_u(\sigma^0\bar\sigma^0)\big)\Big]\,.
\label{eq:49}
\end{equation}
and differs from $Q_{\eta_c}$ given in equation (4) of
\cite{0264-9381-1-1-005} by the explicitly $u$-dependent term of the
second line.  It thus has a similar structure to Penrose's angular
momentum as described in equations (11), (12), and (17a) of
\cite{0264-9381-1-1-005} in the sense that it also differs by a
specific amount of linear supermomentum, but the amount is different
and explicitly $u$-dependent,
$
  Q_{\cY,0,0}= Q^{u=0}_{\cY,0,0}+\half u Q_{0,0,\eth\cY}\,.
$

 The main result derived in \cite{Barnich:2011mi} states that 
 if one is allowed to integrate by parts, and if one defines the ``Dirac bracket'' through
$
   \label{eq:45}
   \{Q_{s_1},Q_{s_2}\}^*[\cX]=-\delta_{s_2}
   Q_{s_1}[\cX]+\Theta_{s_2}[-\delta_{s_1}\cX,\cX], 
$
then the charges define a representation of the $\mathfrak{bms}_4$
algebra, up to a field dependent central extension, 
$
  \{Q_{s_1},Q_{s_2}\}^*=Q_{[s_1,s_2]} +K_{s_1,s_2}, 
$
where $K_{s_1,s_2}$ satisfies the generalized cocycle condition 
$
  K_{[s_1,s_2],s_3}-\delta_{s_3} K_{s_1,s_2}+{\rm cyclic} (1,2,3)=0\,.
$
This representation theorem can be verified directly in the present context \cite{Barnich:2011nu}.

To the best of our knowledge, except for the previous analysis in the
BMS gauge, the above representation result does not exist elsewhere in
the literature. 

A major issue in these considerations is whether one uses the globally
well-defined version of the $\mathfrak{bms}_4$ algebra or a local
version which contains the Virasoro algebra and involves an expansion
in terms of Laurent series. The formulas presented above generally
apply to both cases, except for divergences in the charges that appear
in the second case and have to be handled properly. This is discussed
in more details in \cite{Barnich:2011nu}.

\section*{Acknowledgements}
\label{sec:acknowledgements}

\addcontentsline{toc}{section}{Acknowledgments}

The authors thank C\'edric Troessaert for useful discussions.
  G.B.~is Research Director of the Fund for Scientific Research-FNRS
  (Belgium). This work is supported in part by the Belgian Federal
  Science Policy Office through the Interuniversity Attraction Pole
  P6/11, by IISN-Belgium, by ``Communaut\'e fran\c caise de Belgique -
  Actions de Recherche Concert\'ees'' and by Fondecyt Projects
  No.~1085322 and No.~1090753.



\def\cprime{$'$}
\providecommand{\newblock}{}

\end{document}